\documentclass[12pt]{iopart}

\usepackage{iopams}
\usepackage{amssymb}
\usepackage{graphicx}
\usepackage{color,soul}
\begin{document}

\title[]{Entanglement between distant atoms mediated by a hybrid quantum system consisting of superconducting flux qubit and resonators}

\author{E. Afsaneh$^{1}$, M. Bagheri Harouni$^{*1,2}$, M. Jafari$^{3}$}

\address{$^1$Department of Physics, Faculty of Science,
University of Isfahan, Hezar Jerib St. Isfahan
81764-73441, Iran}
\address{$^2$Department of Physics, Quantum optics group,
University of Isfahan, Hezar Jerib St. Isfahan 81764-73441, Iran}
\address{$^3$Department of Physics, Faculty of Science, International University of Imam
Khomeini-Qazvin, Iran}
\ead{m-bagheri@phys.ui.ac.ir}

\vspace{10pt}
\begin{indented}
\item[]October 2017
\end{indented}

\begin{abstract}
A hybrid quantum system consisting of spatially
separated two-level atoms is studied. Two atoms do not interact directly,
but they are coupled via an intermediate system which is consisting of a superconducting flux qubit interacting
with a mechanical and an electrical resonator which are coupled
to one of the atoms. Moreover, the superconducting flux qubit is
driven by a classical microwave field. Applying the adiabatic
elimination an effective Hamiltonian for the atomic subsystem is
obtained. Our results demonstrate that the
entanglement degradation decay as well as the fidelity decay in the dispersive regime
are faster. Moreover, the driven field
amplitude possesses an important role in the entanglement and
fidelity evolution.
\end{abstract}

\pacs{78.67.Hc, 42.50.Pq, 03.65.Yz, 42.50.Ar}
%
\vspace{2pc}
\noindent{\it Keywords}: Quantum Entanglement, Fidelity, Circuit electrodynamics
%
%
%
%

\section{Introduction}
Coupling between two distant qubits plays an important role in
implementations of quantum information protocols. Mediated features
between distant qubits due to the long-range indirect interaction
provide long coherence times and also can be considered as promising
candidates for quantum state transfer control. Two far coupled atoms can
be entangled and controlled by the photon numbers of thermal field.
One of the two atoms would interact with the thermal field inside the cavity,
while the other would move outside freely \cite{Bashkirov}. In
semiclassical consideration, two distant quantum dipole emitters
in proximity of a metal nanoparticle are entangled by exhibition of
the localized surface plasmons. The steady-state degree of
entanglement only depends on the ratio of distances between the metal nano-particle
and the quantum dipole emitters \cite{Nerkararyan}. Quantum entanglement of two quantum
dot heavy-hole spins separating with a long distance is studied by
using the single-photon interference. Moreover, the heavy-hole spins
demonstrate a long coherence time \cite{Delteil}. Long-distance
coupling between the spin-qubits can be achieved by applying a
long-range interaction such as the qubit coupling to a
ferromagnet \cite{Trifunovic-1} or an electromagnetic field with
significant photon modes \cite{Burkard-1}. Two long distance resonant
exchange qubits are coupled by means of the electromagnetic field in
a microwave cavity. The energy levels of qubits are matched by
the resonant frequency of the cavity \cite{Russ}. Long-range interaction
of spin-orbit coupling is used to prepare the entangled spin qubits
in quantum dots in order to propose a quantum computer
architecture \cite{Trifunovic-2}. Charge transport behavior through
triple quantum dot arrays exhibits long-distance coherent tunnel
coupling between the outward dots \cite{Braakman}. In a hybrid solid
structure, the transfer of quantum information through the
nitrogen-vacancy ensemble acts as the long-distance memory
ingredients which are coupled with the LC circuit as the
transmitter \cite{Zhang}. In all of the mentioned researches, the
distant entangled qubits are resources of quantum information.
Moreover, the state transfer between the distant qubits is an
important step in the quantum information protocols. Therefore, the
introduction and realization of physical systems consisting the
coupling between distant qubits are important steps for the quantum
information processing.\\
To investigate the novel nonclassical quantum information
processing, circuit quantum electrodynamics (circuit QED) and
nanomechanical resonators recently have attracted a great deal of
attention. 

Circuit quantum electrodynamics(cQED) acts as an on-chip analogy of cavity quantum electrodynamics(CQED) at microwave frequencies. 
The circuit QED consists of a superconducting qubit as an artificial atom coupled with microwave resonators. 
Superconducting qubits with dissipationless nonlinearity of the Josephson junction when are connected to microwave resonators provide strong coupling in order to transport single-photons only with classical microwave fields. 

Superconducting devices provide the long
coherence time without dissipation and play a significant role in these
circuits. Superconducting qubits which are interact with a superconducting microwave
resonator were introduced as the circuit QED \cite{Wallraff}. 
In the following, the circuit QED has
progressed considerably in quantum computation \cite{Circuit-QED-1,
Circuit-QED-2, Circuit-QED-3}. For instance, the state transfer
between an electromagnetic resonator and a mechanical resonator was
investigated in the plenty of studies \cite{state transfer-1, state
transfer-2, state transfer-3, state transfer-4}. \\
A system consisting of two superconducting qubits which are coupled
with two or more resonators provides a platform for the quantum
systems in $GHz$ range \cite{two SC qubit-1, two SC qubit-2, two SC
qubit-3, two SC qubit-4}. Two superconducting resonators in
interaction with a superconducting qubit constitute a quantum switch
\cite{two superconducting resonators}. This system is used to
connect two nonlinear microwave resonators \cite{microwave
resonators}. The realization of coupling between a superconducting
flux qubit and two electrical and mechanical resonators has been
demonstrated \cite{Xue}. The entangled superconducting qubits
in a multi-cavity system are studied recently \cite{qubit entanglement}.
Therefore, the superconducting qubits and resonators are one of important
systems in the quantum regime.\\

In CQED, the coupled qubit-cavity systems and also the coupled qubit-spin system have been extensively studied experimentally and theoretically. 
A single quantum system of atom or ion with discrete energy levels is coupled to the quantised radiation field in a cavity in order to developing the single-photon emission\cite{Shore}. 
A coupled atom-cavity system with strong damping which can not follow the strong coupling is considered to investigate the phase response of the system in the regime of high atomic phase-shift experimentally\cite{Schaffer}.
In a two-coupled qubits mediated with a resonator, the energy of a single photon excites two qubits simultaneously. This QED system containing longitudinal couplings operates around the resonant regime\cite{Wang}. The dynamics of qubits-cavity coupled system is studied for quantum states protection, storage and engineering. In this system which is inhomogeneously broadened spin, the states are weakly coupled to the light\cite{Zhukov}.
  
Coupled-spin qubit systems with localized electron spins demonstrate weak inter-qubit couplings which cause low fidelity and reproducibility\cite{Shulman, Veldhorst}. To realize the behavior of the coupled-spin qubit system dynamics, the inter-qubit coupling and environmental noise has been studied theoretically\cite{Sarma}.
In coupling of qubits with optical cavities, the Zeeman splitting of QDs have small magnitude compared with linewidths of cavity which is important in spin–cavity interactions\cite{Carter, Lagoudakis}
In order to enhance the small ground-state splitting, the coupling of QD molecule to cavity is used to achieve the large spin splitting\cite{Vora}.

In comparison the cQED and CQED systems, the advantage of hybrid quantum systems is that the artificial atom have larger transition dipoles than the natural atoms which can enhance the coupling strength magnitude\cite{Blais}. In addition, one-dimensional superconducting microwave resonators contain larger strength of coupling than the ordinary three-dimensional one. These elements can induce strong coupling to the coupled terms through the circuit\cite{Haroche}. These properties of cQED systems explore the quantum optics studies and quantum information researches from microscopic range of atoms to macroscopic of artificial atoms on a chip. Also, the coupling strength of systems in cQED can be tuned easily by manipulating the circuit parameters. In other words, the hybrid systems of cQED show significantly more tunability, scalability and large coupling than the microscopic systems of trapped atom and spin\cite{Wesenberg}. 

In the present contribution, we introduce a physical system which
provides two distant interacting atoms. Our system is based on the
circuit QED setup which is composed of an intermediate superconducting flux
qubit interacting with an electrical resonator as well as a
mechanical resonator. Moreover, each of these resonators are coupled
to a separate two-level atom. The three-level superconducting
flux qubit is driven by a classical microwave field which is
considered as a tuning factor. The two qubits (two atoms) which are coupled indirectly,
provide a platform to achieve the entanglement.
Since this system gives two interaction regimes, resonant and
dispersive, the dynamics of the entanglement and state transfer is
investigated in both interaction regimes. To quantify the
entanglement, a specific measure is required for each quantum
system. The appropriate measure of the entanglement corresponding to the
two-qubit systems is concurrence \cite{concurrence-1, concurrence-2}.
In order to study the state transfer, we would investigate the time
evolution of fidelity. This quantity is defined to know how well
entanglement preserves between the initial entangled state and the
desired state. This concept would characterize the maximal overlap of a
desired state with a maximally entangled state \cite{Jozsa,
fidelity-2, fidelity-3, fidelity-4, fidelity-5}.
\\
This paper is structured as follows. In Sec. 2 , first we
introduce the Hamiltonian of the system. In
the last part of this section, we derive the Markovian master
equation for each case. Next, the evolution of entanglement is studied in Sec. 3 and
the dynamics of fidelity is investigated in Sec. 4 for both resonant and dispersive regimes. In Sec. 4, we obtain the
effective Hamiltonians for both resonant and dispersive regimes as an appendix.
\section{Physical model} \label{Model}
\begin{figure}[h]
\centering
\includegraphics[scale=0.5]{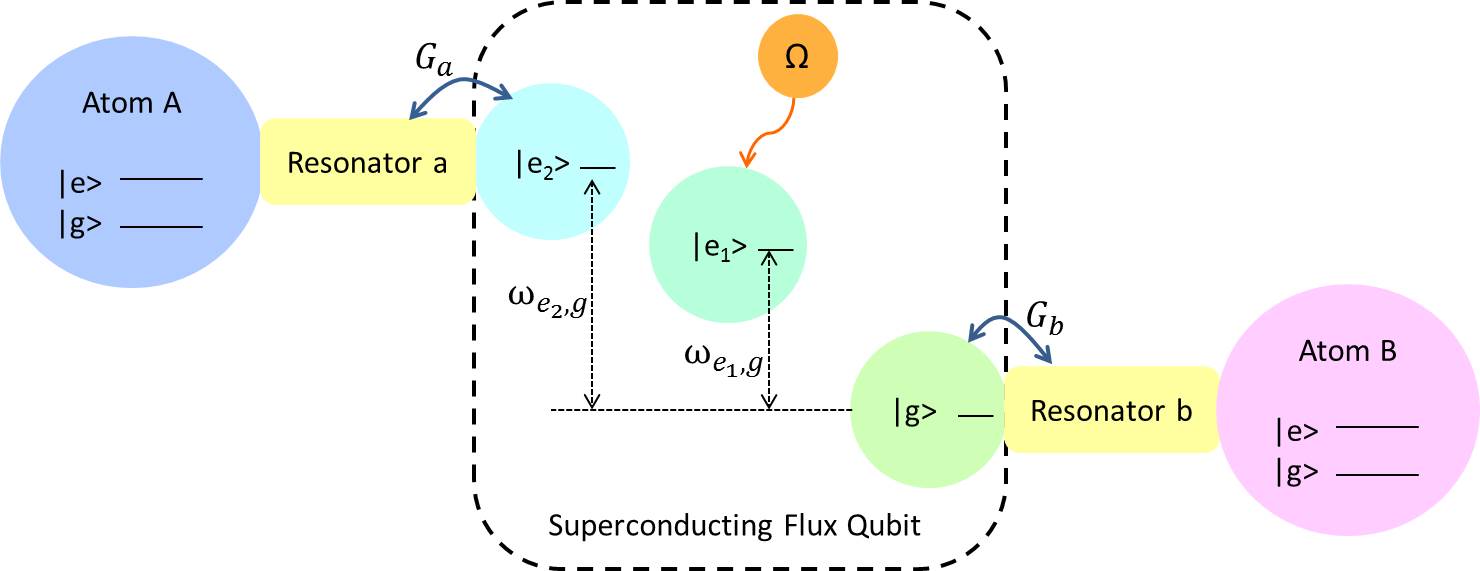}
\caption{The physical system under consideration. Two atoms $A$
and $B$ are coupled indirectly via two resonators $a$ and $b$ which
are interact with a three-level superconducting flux qubit.
The coupling coefficient between second excited flux qubit
level with resonator $a$ is selected as $G_a$ and the coupling strength between the ground state level with
resonator $b$ is chosen as $G_b$. The energy difference between flux qubit
levels are $\omega_{e_2,g}=\omega_{e_2}-\omega_g$ and
$\omega_{e_1,g}=\omega_{e_1}-\omega_g$.The system is
driven by a classical driven field with Rabi frequency $\Omega$.}\label{Picture7}
\end{figure}

The physical system under study is shown in Fig.(\ref{Picture7})
schematically. The proposed system consists of a superconducting
flux qubit which interacts with two resonators, a mechanical
resonator $a$ and an electrical resonator $b$. The flux qubit is
driven by an external classical field with Rabi frequency $\Omega$. Moreover, each resonator is
coupled to a two-level atom, $A$ or $B$. Our purpose is to provide an
effective indirect interaction between the atomic subsystems in
which the driven field, superconducting flux qubit and resonators play the role of a transducer. \\
Usually the coupling strength in the circuit electromechanics is in the
range of the microwave. Therefore, the circuit electromechanics is known as the
microwave counterpart of the cavity optomechanics \cite{Palomaki,
Bagci}. In order to increase the coupling strength of these systems,
it is shown that the replacement of the coupling capacitor by a
superconducting qubit substantially enhances the coupling
strength \cite{Pirkkalainen}. Superconducting qubits are composed of
Josephson junctions which provide the quantum mechanics platform in
macroscopic circuits at low temperatures \cite{Makhlin}.\\
The nonlinear nature of Josephson junction leads to a nonuniform
separation of energy levels \cite{You}. Taking into account two
lowest energy levels, this system would be supposed as a qubit.
Superconducting qubits are divided into three categories: charge, flux and
phase qubit. In the flux qubit, the Josephson coupling energy is greater
than the Coulomb energy which can be controlled by an external
microwave field. This applied microwave field would be used to
manipulate the flux qubit \cite{Nakamura}.\\
Resonators are considered as devices which are able to carry the
electromagnetic field and to exchange energy between two ultimate
destinations, such as atoms, through interaction with them.
Resonators with high quality factor would work in $GHz$ frequency
regime and would be divided into superconducting and nanomechanical
resonators \cite{Wallraff, Cleland}. One kind of superconducting
resonators is the LC resonator which utilizes a tunable electrical
element. This type of resonator is known as an
electrical resonator.
\subsection{Hamiltonian}
The Hamiltonian of the whole system may be written as
$H=H_0+H_{int}$. In this relation, $H_0$ corresponds to the free
Hamiltonian and is composed of the following terms:
\begin{equation}\label{H0}
H_0=H_A+H_R+H_q.
\end{equation}
In this equation, $H_A$ describes the free Hamiltonian of atoms A and B:
\begin{equation}
H_A=\hbar\omega_A\sigma^A_z+\hbar\omega_B\sigma ^B_z.
\end{equation}
The atomic energy transitions are given by $\omega_A$ and
$\omega_B$ and also $\sigma^i_z=|e_i\rangle\langle
e_i|-|g_i\rangle\langle g_i |$ in which $i=A$ or $B$. We have assumed that
$|e_i \rangle(|g_i \rangle)$ is the excited (ground) state of the
$i$th atom. Moreover, in Eq.(\ref{H0}) $H_R$ is the free Hamiltonian of the electrical and
mechanical resonators and is defined as
\begin{equation}
H_R=\hbar\omega _aa^{\dagger}a+\hbar\omega _bb^{\dagger}b.
\end{equation}
In the present system, we have supposed that every resonator possesses a
single mode with frequency $\omega_a$ and $\omega_b$. In this
equation, $a$ and $b$ are the annihilation operators associated with
the resonators. The third term of Eq.(\ref{H0}) corresponds to the
free Hamiltonian of the flux qubit. This flux qubit consists of
three Josephson junctions in a loop configuration. For this qubit,
the two lowest energy states are localized while the third one is
delocalized. Biasing the loop by a magnetic flux, this system may
serve as a $V$ type artificial atom with a cyclic transition
configuration \cite{Xue}. If the three lowest energy levels
of this loop are labeled as $|k\rangle$ with $k\in\{g, e_1, e_2\}$, the qubit
Hamiltonian could be written as
\begin{equation}
H_q=\sum_k\hbar\omega_k \sigma^k_z, \hspace{1cm} k=g, e_1, e_2.
\end{equation}
The interaction Hamiltonian between different subsystems can be
written as
\begin{equation}\label{Hint}
H_{int}=H_{A,a}+H_{B,b}+H_{a,q}+H_{b,q}+H_{Driv}.
\end{equation}
In this equation, $H_{A,a}$ and $H_{B,b}$ describe the interaction
between the resonators, $a$ and $b$, with the two-level atoms, $A$
and $B$. In the present contribution, we are going to study two different
resonant and dispersive interaction regimes. The interaction between
atoms and resonators would be described with the following
Hamiltonians \cite{Bourassa}:
\begin{equation}\label{HAa}
H_{A,a}= G_A(a^{\dagger}\sigma_A^{-} +a\sigma_A^{\dagger}), \, \, \, H_{B,b}=G_B(b^{\dagger}\sigma_B^- +b\sigma_B^{\dagger}).
\end{equation}
In this equation, $\sigma_A^{-}=|g_A\rangle\langle e_A|$ and
$\sigma_B^{-}=|g_B \rangle \langle e_B| $ are the lowering operators of atomic systems.
In these relations, $G_A$ and $G_B$ are the coupling strengths between the resonators
and two-level atoms. Therefore, the transition frequency of the atom $A$
(qubit $A$), $\omega_A$, should be close to the frequency of electrical resonator
$a$, $\omega_a$, and also $\omega_B$, the transition frequency of atom $B$ (qubit $B$)
must be near to $\omega_b$, the frequency of mechanical resonator $b$.
\\
Energy scale of each physical system determines the physical regime of the system.
Also, the energy scale determines which physical system could be coupled to each other.
Therefore, an important point about the hybrid systems corresponds to their energy scales.
For a typical qubit loop, the energy difference
between qubit eigenstates may be within the range of $2\pi \times
[0,10] GHz$ \cite{Paauw}. Furthermore, the electrical resonator can
be assumed as an $LC$ part of a superconducting transmission line
forming a one dimensional cavity with frequency regime $\sim 2\pi
\times[1,10]GHz$ \cite{Devoret}. Moreover, the recent progress of experimental
techniques reveal that the mechanical resonators would be characterized in
$GHz$ frequency regimes \cite{OConnell}.
\\
In the same manner, the $H_{a,q}$ and $H_{b,q}$ terms in Hamiltonian (\ref{Hint}) describe
the interaction between the resonators and artificial atom (qubit subsystem). In the qubit
eigen basis, these coupling are described by
\begin{equation}
H_{a,q}=G_a (a^{\dagger}\sigma_{e_2,g}^- +a \sigma_{e_2,g}^+), \, \, \, H_{b,q}=G_b (b^{\dagger} \sigma_{e_1,g}^- +b \sigma_{e_1,g}^+).
\end{equation}
In these relations, $G_a$ and $G_b$ are the coupling strengths
between resonators and qubit systems. Also,
$\sigma^{-}_{x,y}=|x\rangle\langle y|$ is a lowering operator.
As the resonator $a$ is coupled with $e_2\leftrightarrow g$ transition of the
superconducting qubit, their frequencies need to be closed to each other
similar to the resonator $b$ and $e_1\leftrightarrow g$ transition of superconducting qubit.
In a recent experiment the coupling strengths between the electrical resonator and
flux qubit as well as the mechanical resonator are obtained of the order of $\mathcal{O}(1) MHz$ \cite{OConnell}-\cite{Abdumalikov}.
Therefore, we choose the coupling strength of all coupled subsystems as the same $ G_a=G_b=G_A=G_B=40MHz$.
\\
In Hamiltonian(\ref{Hint}), $H_{Driv}$ is the driven interaction of
the flux qubit. In order to manipulate the system, the flux qubit
should be inductively driven by an external microwave field. We have
supposed that this driven field dispersively couples with the
transition of $|e_1\rangle\leftrightarrow|e_2\rangle$. This driven
interaction would be modeled as
\begin{equation}
H_{Driv}=  \Omega (\sigma^{-}_{e_1,e_2}+\sigma^{+}_{e_1,e_2}).
\end{equation}
Here, $\Omega$ is the coupling coefficient of the microwave driven
field and is related to the amplitude of the driven field. This
quantity possesses an important role in the dynamics of the present
system. The system under consideration is composed of different
subsystems and its evolution provides a complex dynamics.
\section{Dynamics of system}
The main physical system is a hybrid system which is composed of two atoms,
the resonators and one flux qubit which is driven by a classical field.
The atomic decay rate is of the order of
$GHz$. In comparison to this decay rate, the electrical resonator
decay rate is of several $kHz$, whereas the mechanical resonators
possess the usual decay rates of several $MHz$. Furthermore, an
interesting feature of superconducting qubits is their long
decoherence times \cite{Burkard-2}. 


To study the dynamics of the present system, we start from the Liouville-von Neumann equation for the complete system in the interaction picture\cite{Breuer}. The whole system consists of a multilevel structure involves two indirectly coupled qubits $A$ and  $B$ mediated by an electrical resonator $a$ and a mechanical resonator $b$ connecting to a superconducting flux qubit. To simplify the computational process and obtain the effective Hamiltonian, the procedure of adiabatic elimination is applied\cite{Walls,Shore-2,Yoo}. Superconducting flux qubit with three-level lambda type system is reduced to a two-level one with adiabatic elimination strategy\cite{Allen,Brion,Azouit}.
Also, for single-photon transport a superconducting transmission line resonator (TLR) array coupled with a Cooper pair box(CPB) was used in order to connect two TLRs.  In this research, an effective interaction between the TLRs was obtained by adiabatically eliminating the variables of the CPB\cite{Liao}.
By adiabatic elimination of atomic and photonic states, the coupling of atomic qubits at a quantum network with some cavities coupled with optical fibers is performed leading to qubit-qubit interactions\cite{Zheng}.
In another study, by adiabatic elimination of atomic and photonic states atomic qubits at a quantum network with some cavities which are coupled to optical fibers lead to qubit-qubit interactions\cite{Zheng-2}.
Therefore, we can define an effective Hamiltonian for our system by adiabatic elimination of superconducting flux qubit with three-level to a two-level one firstly and then by adiabatically eliminating of mediated resonators. The effective Hamiltonian is calculated for dispersive and resonant regimes in Appendix. Our effective open system of two distant qubits is coupled to the common fermionic reservoir. To obtain the time evolution of the system, we trace out the bath degrees of freedom which defines the lead correlation function of master equation. Then under the Born-Markov approximations, we calculate the quantum master equation(QME) for the reduced density matrix which is obtained: 
\begin{equation}
\dot{\rho}(t)=-\frac{i}{\hbar}[H_{I},\rho(t)]+\mathcal{L}_{A} \rho+\mathcal{L}_{B} \rho.
\end{equation}
where $\rho$ denotes the reduced density matrix of system. The first term shows the lamb shift and $ \mathcal{L}_i\rho=\frac{\Gamma_{ij}}{2}[2\sigma_{i}^{-} \rho \sigma_{j}^{+} -\sigma_{j}^{+} \sigma_{i}^{-}\rho-\rho\sigma_{j}^{+} \sigma_{i}^{-}]$, $i, j=A, B $ is the Lindblad operator which describes the dissipation in the system. It is worth to note that the introduced hybrid system leads to an indirect interaction between two remote atoms. Thus, we have supposed that the effective physical system is composed of two interacting atoms. So, the master equation for the present system is achieved:
\begin{eqnarray} \label{masterEq}
\dot{\rho}&= &- iG[\sigma_{A}^{-} \sigma_{B}^{+} \rho +\sigma_A^+ \sigma_B^- \rho-\rho \sigma_A^- \sigma_B^+ - \rho \sigma_A^+ \sigma_B^-] \nonumber\\
&+&\frac{\Gamma_{A}}{2}[2\sigma_A^- \rho \sigma_A^+ -\sigma_A^+ \sigma_A^-\rho-\rho\sigma_A^+ \sigma_A^-]
+\frac{\Gamma_{B}}{2}[2\sigma_B^- \rho \sigma_B^+ -\sigma_B^+ \sigma_B^-\rho-\rho\sigma_B^+ \sigma_B^-]\nonumber\\
&+&\frac{\Gamma_{AB}}{2}[2\sigma_A^- \rho \sigma_B^+ -\sigma_B^+
\sigma_A^-\rho -\rho\sigma_B^+ \sigma_A^-]
+\frac{\Gamma_{BA}}{2}[2\sigma_B^- \rho \sigma_A^+ -\sigma_A^+
\sigma_B^-\rho -\rho\sigma_A^+ \sigma_B^- ].
 \end{eqnarray}
In this equation, the dissipation coefficient $\Gamma_{ij}$($i, j=A, B $) denotes the effective atomic decay rates without loss of generality relating to the bath correlation function which is defines:
\begin{equation}\label{dissipation}
\Gamma_{ij}(\omega)=  2 \pi \sum_{i \nu \sigma} t_{i \nu \sigma} t^{*}_{j \nu^{'} \sigma^{'}} \langle  c^{\dagger}_{\nu} c_{\nu} \rangle_{Bath}
\end{equation}
in which, $c(c^{\dagger})$ indicates the annihilation(creation) fermionic operator of reservoir, $\nu$ shows the wave vector and $\sigma$ denotes the spin of the central system. Here, we assume $\Gamma_{A}=\Gamma_{B}=2\pi\times0.1MHz$. Also, as the qubits do not interact with each other directly, the induced indirect interaction is absent so we have $\Gamma_{AB}=\Gamma_{BA}=0$.

In the present contribution, we study
the system in the computational basis, that is $ |1 \rangle =
|ee\rangle$, $ |2\rangle=|eg\rangle$, $|3\rangle=|ge\rangle$ and
$|4\rangle =|gg\rangle$. Here, $|ij\rangle=|i\rangle_A |j\rangle_B$
are the excited or ground states of atomic systems. As indicated
above, the introduced hybrid system provides an effective
interaction between two atoms. Therefore, in this system we
encountered with two coupled qubits. An interesting physical quantity
would be the entanglement between qubits (atomic subsystems). To
quantify the entanglement in the two-qubit systems, concurrence is an
appropriate measure for both pure and mixed
states \cite{concurrence-1,concurrence-2}. Accordingly, we will
obtain the concurrence in the present system.
\subsection{Concurrence}\label{Concurrence}
\begin{figure}[t]
\centering
\includegraphics[width=0.48\columnwidth]{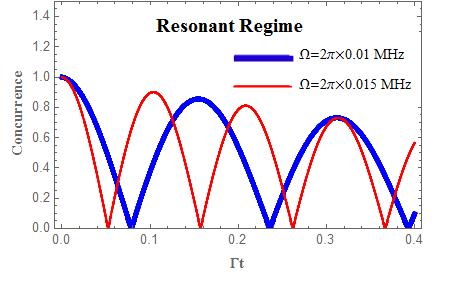}
\hspace{0.1cm}
\includegraphics[width=0.48\columnwidth]{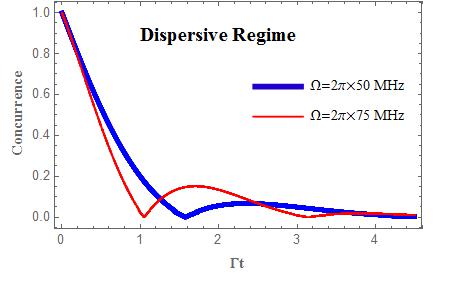}
\caption{Left panel: concurrence for resonant regime, thick line $\Omega_R=2\pi\times0.01MHz$, thin line $\Omega_R=2\pi\times0.015MHz$. Right panel: concurrence for dispersive regime, thick line $\Omega_D=2\pi\times50MHz$, thin line $\Omega_D=2\pi\times75MHz$.}\label{Concurrence-R-D}
\end{figure}

\begin{figure}[h]
\centering
\includegraphics[scale=0.78]{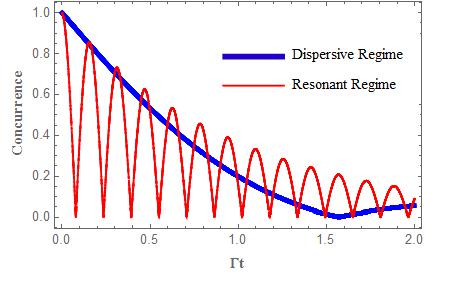}
\caption{Comparison of Concurrence between resonant and dispersive regimes. Thick line: dispersive regime $\Omega_D=2\pi\times50MHz$ ; thin line: resonant regime $\Omega_R=2\pi\times0.01MHz$.}\label{Concurrence-Comparison-10}
\end{figure}

Concurrence is defined as $C(\rho)=Max [
0,\lambda_1-\lambda_2-\lambda_3-\lambda_4 ] $, where
$\lambda_i,(i=1,2,3,4)$ are non-negative eigenvalues of a matrix $R$
with decreasing order $\lambda_1>\lambda_2>\lambda_3>\lambda_4$. The
matrix $R$ is defined as $ R=\sqrt{\sqrt{\rho}\tilde{\rho}\sqrt{\rho}} $, where $
\rho$ refers to the density matrix of the system and
$\tilde{\rho}=(\sigma_y \otimes \sigma_y)\rho^*(\sigma_y \otimes
\sigma_y)$. Here, $\sigma_y$ represents the $y$ component of the
Pauli matrices and $\rho^* $ is the complex conjugate of the density
matrix. The time-dependence of the concurrence in both interaction
regimes, resonant and dispersive, is shown in
Fig.(\ref{Concurrence-R-D}).
In this figure each panel corresponds to the specific interaction
regime, whereas in each case different plots correspond to a given
driven field amplitude, $\Omega$. The driven field is an
experimental parameter to control the dynamics of the system. The
relevant deserved magnitude for a given field depends on the validity
of adiabatic elimination. This validity which leads to the effective
Hamiltonian by elimination of irrelevant states, is that the driven
field Rabi frequency should be less than detuning considerably \cite{Brion}.
 \\
The initial state of atoms is considered as
\begin{equation}\label{Bell}
\rho[0]=
\left[ {\begin{array}{cccc}
    0 & 0 & 0 & 0 \\
    0 & 0.5 & -0.5i & 0  \\
    0 & 0.5 i & 0.5 & 0 \\
    0 & 0 & 0 & 0 \\
  \end{array} } \right],
\end{equation}
which possesses the highest degree of entanglement.
The interesting poit about the present system is that during the evolution, the qubits can not be entangled for the initial separable state. 
The main reason of this event is that when the qubits from the first are in interaction with each other and then they experience an additional interaction which is different for each of them, the qubits would be entangled even they start from separable initial state.
The figures show the
entanglement degradation of the system as time elapses. Additionally, the
entanglement degradation is governed by the driven field.
Therefore, the classical microwave driven field may be used as a
control parameter for the entanglement between remote atoms. A comparison
between the panels of Fig.(\ref{Concurrence-R-D}) demonstrates that the
entanglement between atoms follows an oscillatory time evolution. In the resonant regime the period is smaller than the dispersive regime. Moreover, the entanglement degradation in the dispersive regime is faster than resonant one. This comparison between two interaction regimes is illustrated in
Fig.(\ref{Concurrence-Comparison-10}).
This figure exhibits long lasting entanglement in the resonant
regime. That is, the entanglement degradation is faster in the
dispersive regime. As a consequence, the interaction
between atoms and resonators provides a platform for the entanglement in
the present system. Additionally, the driven field is another
important parameter in the entanglement time evolution.
\subsection{Fidelity}\label{Fidelity}
\begin{figure}[t]
\centering
\includegraphics[width=0.48\columnwidth]{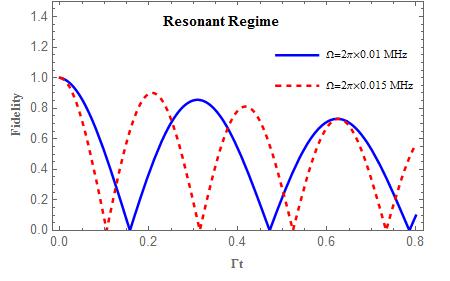}
\hspace{0.1cm}
\includegraphics[width=0.48\columnwidth]{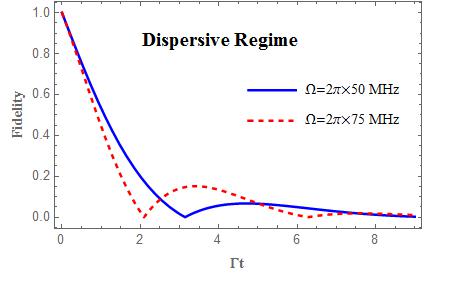}
\caption{Left panel: fidelity for resonant regime, solid line $\Omega_R=2\pi\times0.01MHz$ and dashed line $\Omega_R=2\pi\times0.015MHz$. Right panel: fidelity for dispersive regime, solid line $\Omega_D=2\pi\times50MHz$, dashed line $\Omega_D=2\pi\times75MHz$. }\label{Fidelity-R-D}
\end{figure}

\begin{figure}[h]
\centering
\includegraphics[scale=0.78]{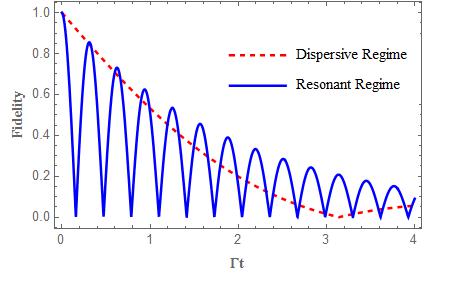}
\caption{ Comparison of Fidelity between resonant and dispersive regimes. Solid line: resonant regime,  $\Omega_R=2\pi\times0.01MHz$; dashed line: dispersive regime, $\Omega_D=2\pi\times50MHz$.}\label{Fidelity-Comparison-10}
\end{figure}

The physical system under study can be used for state transfer
between atoms. In this situation, fidelity is an important parameter
which characterizes the overlap of a desired state with a maximally
entangled one. In the present contribution, to quantify the fidelity
we have chosen the Bell state which is used for the concurrence as the reference state. The fidelity of a mixed
state is defined as  $F(\rho_1,\rho_2)=max[\langle \Phi_1| \Phi_2
\rangle|^2]$ \cite{Jozsa}, which is known as Uhlmann
formula \cite{Uhlmann}. In this relation, $|\Phi_1\rangle$ and
$|\Phi_2 \rangle$ are the purifications of $\rho_1$ and $\rho_2$,
respectively. This expression of fidelity can be written in an
equivalent relation $F(\rho_1,\rho_2) =Tr[\sqrt{\sqrt{\rho_1}\rho_2
\sqrt{\rho_1}}] $. In this definition, $\rho_1$ is the initial
state, or target state, and $\rho_2$ is the elapsed state. Then,
this quantity determines to what extent the evolved state of the
system is close to the target state $\rho_1$. The fidelity time evolution
in the system under study is illustrated in
Fig.(\ref{Fidelity-R-D}) for both the resonant and dispersive regimes.
Different plots in each panel correspond to the given driven field.
Here, our target state is a Bell state and we choose the Bell state which is given in Eq. (\ref{Bell}) as the initial
state of the system.
For that, the fidelity started
at the maximum value and as time elapses it is decayed to zero. That is,
the initial maximally entangled state evolves into a complete
separable final state. In other words, the fidelity follows the similar
scenario as the entanglement. Moreover, a comparison between the
time evolution of the fidelity in two different resonant and dispersive
regimes is illustrated in Fig.(\ref{Fidelity-Comparison-10}).
This plot shows that the fidelity decay rate is faster in the
dispersive interaction regime, similar to the entanglement evolution.
Thus, the resonant interaction between atoms and resonators keeps
the atomic subsystem state closer to the Bell state.
\section{Conclusion}\label{Conclusion}
We have proposed an analysis on an electromechanical circuit consists
of a superconducting flux qubit which interacts with the electrical
and mechanical resonators. In turn, each resonator is in interaction
with a single two-level atomic system. The flux qubit is driven by a
microwave field. We have shown that in the specific situation, one
could achieve an effective interaction between two atoms in the
system. Therefore, the introduced hybrid system reduces to a
coupled-qubits system. Next, the time evolution of the quantum entanglement
and fidelity were studied. The interaction between atoms and
resonators is considered in two different resonant and dispersive
regimes. Our results have illustrated that the entanglement degradation in the
dispersive interaction is faster than the resonant interaction regime.
Similarly, the decay rate of the fidelity in the dispersive regime is faster
than the resonant interaction regime. That is, the entanglement and fidelity
follow the similar time evolution in two interaction regimes. Also, the increase of the driven field
amplitude leads to slower entanglement degradation as well as
fidelity decay. As a result, in the introduced system the resonant
interaction regime and the strong driven field lead to more robust
entanglement. Moreover, the dispersive interaction between the atoms
and resonators provides an appropriate platform for the state transfer.
\section*{Appendix: Calculation of Dispersive and Resonant Regime} \label{appendix} 
 To obtain the effective Hamiltonian in dispersive and resonant regimes, all parts of the total Hamiltonian should be written in interaction picture:
\begin{eqnarray}
\tilde{H}_I(t)&= & G_A(\hat{a}^{\dagger}\hat{\sigma}_A^{-} e^{i\Delta_{Aa}t}+\hat{a}\hat{\sigma}_A^{+} e^{-i\Delta_{Aa}t})
+G_a(\hat{a}^{\dagger}\hat{\sigma}_{ge_{2}}^{-} e^{i\Delta_{aq}t}+\hat{a}\hat{\sigma}_{ge_{2}}^{+} e^{-i\Delta_{aq}t})  \nonumber\\
&+ &  G_B(\hat{b}^{\dagger}\hat{\sigma}_B^{-} e^{i\Delta_{Bb}t}+\hat{b}\hat{\sigma}_B^{+} e^{-i\Delta_{Bb}t})
+  G_b(\hat{b}^{\dagger} \hat{\sigma}_{ge_{1}}^{-} e^{i\Delta_{bq}t}+\hat{b}\hat{\sigma}_{ge_{1}}^+ e^{-i\Delta_{bq}t})  \nonumber\\
&+ & \Omega(\hat{\sigma}_{e_{1}e_{2}}e^{i\Delta_{e_{1}e_{2}}t}+\hat{\sigma}_{e_{1}e_{2}}^+e^{-i\Delta_{e_{1}e_{2}}t}).
\end{eqnarray}
Here, $G_a$, $G_b$, $G_A$ and $G_B$ denote the coupling strength of respectively and $\Omega$ indicates the frequency of the external driven field. Also we describe the detuning parameters as: $\Delta_{Aa}=|\omega_a-\omega_A| $, $\Delta_{aq}=|\omega_{ge_{2}}-\omega_a|$, $\Delta_{Bb}=|\omega_b-\omega_B|$, $\Delta_{bq}=|\omega_{ge_{1}}-\omega_b|$ and $\Delta_{e_{1}e_{2}}=\omega_{ge_{2}}-\omega_{ge_{1}}$. 
The electrical and mechanical resonators work in the order of $GHz$ frequency, we typically consider their frequencies as $\omega_a=2\pi \times 5.12 GHz$ and $\omega_b=2\pi \times 1.4 GHz$ respectively\cite{Xue}. 
Resonators are coupled to the qubits $A$ and $B$ in one side and superconducting flux qubit in other side, so the frequency of each coupled qubit should be in the range of the mutual resonator. 
The level of qubits can be tuned by means of the gate voltage biasing, external applied microwave or magnetic flux which let us to set the frequency of the subsystems for each regimes.

In the following, we calculate the effective Hamiltonian for the dispersive and resonant regimes.
\subsection{Dispersive Regime}
In the dispersive regime, the coupled resonator-atom subsystems are far detuned comparing with their coupling strength. This limitation for the present system can be  expressed as $\Delta_{aq} \gg G_a$, $\Delta_{bq} \gg G_b$, $\Delta_{Aa} \gg G_A$ and $\Delta_{Bb} \gg G_B$.
In addition, we suppose that $\Delta_{aq} \Delta_{bq} \gg \Omega_D^2$. 
However, in this regime, the interaction does not lead to the energy exchange and the interaction effects would be followed through frequency shift and other physical phenomena as well\cite{Boissonneault}.
\\
According to the constant magnitude of coupling strength for each part and also the conditions of disspersive regime, we arrange the detuning parameters $\Delta_{aq}=\Delta_{bq}=\Delta_{Aa}=\Delta_{Bb}=2\pi \times 120 MHz$. 
As the resonators work in the order of $GHz$ frequency, we consider the frequency of the electrical and mechanical resonators as $\omega_a=2\pi \times 5.12 GHz$ and $\omega_b=2\pi \times 1.4 GHz$ respectively \cite{Xue}. With respect to the coupling of the qubit $A$ and the $e_2,g$ level of superconducting qubit with resonator $a$ from both sides, the relevant frequencies are choosen $\omega_A=2\pi \times 5.24 GHz$ and $\omega_{e_2,g}=2\pi \times 5 GHz$ respectively. Also, we select $\omega_{e_1,g}=2\pi \times 1.52 GHz$ and $\omega_{B}=2\pi \times 1.28 GHz$ for $e_1,g$ level of superconducting qubit and qubit $B$ which are connected to resonator $b$ respectively.

According to the adiabatic elimination, the Hamiltonian for the present regime is transformed as $H^{eff}_R=e^{\lambda_D} H e^{\lambda_D^{\dagger}}$. The new operator $\lambda_D$ is defined as
 \begin{eqnarray}
\lambda_D &=&\frac{G_A}{\Delta_{Aa}}(\hat{\sigma}_{A}^{-} \hat{a}^{\dagger}-\hat{\sigma}_{A}^{+}\hat{a})+
\frac{G_a}{\Delta_{aq}}(\hat{\sigma}_{ge_{2}}^{-} \hat{a}^{\dagger}-\hat{\sigma}_{ge_{2}}^{+}\hat{a}) +
\frac{\Omega}{\Delta_{e_{1}e_{2}}}(\hat{\sigma}_{e_{1}e_{2}}-\hat{\sigma}_{e_{1}e_{2}}^{+} )\nonumber\\
&+& \frac{G_B}{\Delta_{Bb}}(\hat{\sigma}_{B}^{-}\hat{b}^{\dagger}-\hat{\sigma}_B^+\hat{b})+\frac{G_{b}}{\Delta_{bq}}(\hat{\sigma}_{ge_{1}}^{-}
\hat{b}^{\dagger}-\hat{\sigma}_{ge_{1}}^{+}\hat{b}).
 \end{eqnarray}
 Using the Barker-Campbell-Hausdorff relation, the transformed Hamiltonian could be expanded as:
\begin{equation}
H^{eff}_D=H+[\lambda_D,H]+ \frac{1}{2} [\lambda_D, [\lambda_D,H]] + \cdots 
\end{equation}
Following this expansion up to fifth order, the effective Hamiltonian describing the remote interaction between atomic subsystems is  obtained as 
 \begin{equation} \label{HD}
H^{eff}_D=G_D(\hat{\sigma} _{A}^{-} \hat{\sigma} _{B}^{+} +\hat{\sigma} _{A}^{+} \hat{\sigma} _{B}^{-}) 
 \end{equation}
here $G_D$ is an effective coupling strength in the dispersive regime and is given as
\begin{equation}
G_D=\frac{1}{20}\frac{\Omega G_a G_b G_AG_B}{\Delta_{e_1e_2}\Delta_{aq}\Delta_{bq}\Delta_{Aa}\Delta_{Bb}}\Delta_{Aa}-\Delta_{Bb}+4(\Delta_{aq}-\Delta_{bq})+6\Delta_{e_1e_2}],
\end{equation}
\subsection{Resonant Regime}
The resonant regime is defined when the coupled subsystems(qubit-resonator or atom-resonator) are in resonant or near-resonant limit. 
The limitation of this regime in the physical realization is $\Delta_{aq}\ll G_a$, $\Delta_{bq}\ll G_b$, $\Delta_{Aa}\ll G_A$ and $\Delta_{Bb} \ll G_B$.
\\
Therefore in these circumistances, the frequency of qubits should be so close to resonators that we choose them in this way: $\omega_{A}=2\pi \times 5.135 GHz$ and $\omega_{e_2,g}=2\pi \times 5.09 GHz$ which are coupled with resonator $a$, moreover $\omega_{e_1}=2\pi \times 1.43 GHz$ and $\omega_{B}=2\pi \times 1.385 GHz$  which are connected to the resonator $b$. However the frequency of resonators are considered the same as before. 
\\
These consideration provide us to have the same detuning parameters of $\Delta_{aq}=\Delta_{bq}=2 \pi \times 30 MHz$ and $\Delta_{Aa}=\Delta_{Bb}=2 \pi \times 15MHz$.
In addition, we suppose that $\Delta_{aq} \Delta_{bq} \gg \Omega_R^2$. 
To obtain an effective Hamiltonian for our distant qubits system in resonant regime, we apply the time evolution operator\cite{Schleich}:\\
\begin{equation}\label{unitary}
u(t,t_0=0)\cong 1 -\frac{i}{\hbar}\int^{t}_{0} dt^{'} H_{int}(t^{'})-\frac{1}{\hbar^{2}}\int^{t}_{0} dt^{'} H_{int}(t^{'}) \int^{t^{'}}_{0} dt^{''} H_{int}(t^{''})+ \dots
\end{equation}
after calculation, the unitary transformation becomes $u(t,t_0=0)\cong 1 -\frac{i}{\hbar} H_{eff}t$ which can be written as $u(t,t_0=0)\cong exp[-\frac{i}{\hbar} H_{eff}t]$. For our system, we calculate the unitary equation(\ref{unitary}) up to fifth order and avarage the unitary relation for the ground state $|0\rangle$ which becomes:
\begin{equation}
u(t,t_0=0)\cong 1 -\frac{i}{\hbar} \frac{ 2 \Omega G_aG_bG_AG_B}{\Delta_{Aa} \Delta_{Bb}(\Delta_{aq}-\Delta_{Aa})(\Delta_{bq}-\Delta_{Bb}) }
\end{equation}
Therefore, the effective Hamiltonian for resonant regime is achieved:
\begin{equation}\label{HR}
H^{eff}_R=G_R(\hat{\sigma}_B^+ \hat{\sigma}_A^- +\hat{\sigma}_A^+ \hat{\sigma}_B^-)
\end{equation}
in which the coupling coefficient for resonant describes as:
\begin{equation}
G_R=2 \Omega G_aG_bG_AG_B \frac{1}{\Delta_{Aa} \Delta_{Bb}(\Delta_{aq}-\Delta_{Aa})(\Delta_{bq}-\Delta_{Bb}) }
\end{equation}
Here, we use this detuning parameter $\Delta_{Bb}-\Delta_{bq}-\Delta_{e_1e_2}+\Delta_{aq}-\Delta_{Aa}=0 $ in resonant regime.
\\
 Drawing a comparison between Hamiltonians in Eq.(\ref{HD}) and Eq.(\ref{HR}) reflects this fact that the both effective Hamiltonians describe an indirect interaction between atomic subsystems. In this case, the significant point is concerned about the difference order of magnitude between the coupling strengths of these interaction regimes which is $\frac{G_R}{G_D}\mathcal{O} 10$. Therefore, the introduced system may be supposed as a coupler between distanced atomic systems.

\end{document}